\documentclass[conference]{IEEEtran}
\pdfoutput=1
\usepackage[T1]{fontenc}
\usepackage[utf8]{inputenc}
\usepackage{url}
\usepackage{graphicx}
\usepackage{listings}
\usepackage{algorithmic}
\usepackage{flushend}

\begin{document}

 \title{A Visual Programming Language for Drawing and Executing Flowcharts}
 \author{\IEEEauthorblockN{Drazen Lucanin}
 \IEEEauthorblockA{Faculty of Electrical Engineering\\and Computing\\
 University of Zagreb\\
 Email: drazen.lucanin@gmail.com}
 \and
 \IEEEauthorblockN{Ivan Fabek}
 \IEEEauthorblockA{Faculty of Electrical Engineering\\and Computing\\
 University of Zagreb\\
 Email: ivan.fabek@gmail.com}}
 
 \maketitle
 
 
 \begin{abstract}
 With recent advances in graphical user interfaces, more and more tasks on computers have become easier to perform. Out of the belief that creating computer programs can also be one of them, visual programming languages (VPLs) have emerged. The goal of VPLs is to shift a part of work from the programmer to the IDE so that the programmer can focus more on algorithm logic than the syntax of the implementation programming language.In this article, the methods required to build a VPL are presented, with an emphasis on a novel method of code generation in a WHILE language. Also, the methods for achieving basic principles of VPLs will be shown -- suitable visual presentation of information and guiding the programmer in the right direction using constraints.
  
These methods are demonstrated on an example of vIDE, a VPL based on the Eclipse integrated development environment (IDE). The design of vIDE with respect to the Eclipse Graphical Modeling Framework (GMF) is described. The concept of a flowchart graphical notation is examined in contrast with the algorithm model it maps to. Finally, the disambiguity of the model representation of an algorithm is discussed and the methods for transforming it to an actual implementation in a programming language.

 \end{abstract}
 
 
 \section{ Introduction}

\IEEEPARstart{A}{s} desktop computers become more and more powerful, we see a big improvement in the quality of graphical user interfaces (GUIs). What about programming? Can the classical textual programming language environments be replaced with more modern graphical applications? There are few key points in discussing positive aspects of developing VPLs.

\subsection{The strengths of visual programming}
\label{sec:strengths}

According to its definition in \cite{knuth_art_1997}, an algorithm has five important features: \textit{finiteness, definiteness, input, output and effectiveness}. The most important feature we will examine in this context is definiteness -- the steps of an algorithm have to be well defined, disambiguous. We can infer from this that to implement an algorithm, a computer program also requires definiteness.

In classical programming languages (such as C, defined in \cite{kernighan_c_1988}) the programmer writes his program down in a textual file, compiles it to machine code and is able to execute it. Textual files can consist of any imaginable text -- from Shakespeare's dramas to random gibberish. Well defined classical programs are a very small subset of text. In classical programming languages, definiteness is checked mostly through programming language \textit{syntax} (it is also achieved in part through semantic and dynamical checks at later stages of program compiling and running, but that won't be discussed here). Syntax is essentially a set of rules that constraint text that qualifies as valid -- more details can be found in \cite{hopcroft_introduction_2007}. The programmer tries to describe an algorithm he has imagined by writing commands in some syntax and getting notifications about their correctness afterwards. This is in its essence a very black-box approach, i.e. the programmer receives only a posteriori notifications about the program's definiteness. Of course, there are some features that help in the process, like dynamic type checking or auto-completion which save time, but they do not let the programmer fully forget about the syntax.

In VPLs the programmer doesn't input text, but uses modern GUI capabilities to drag \& drop blocks of commands, connections between them etc. to describe an algorithm. This is by itself a reduction of the possible constructs that can be made by a programmer -- only a couple of symbols (defined by the language designer) are available, in contrast to the whole alphabet (and more) in classical programming. \textbf{The choices a programmer has to make when defining programs are limited to the ones that satisfy the language syntax only.} An illustrative way to describe this would be to say that in VPLs the keyboard has been sawed off to contain only the keys vital to programming (sort of like the old Spectrum computers). Another important aspect is \textbf{the chance for preeminent notifications} -- notifying the programmer about the possible constraint breeches (syntactic or semantic errors) while he is still dragging an element in the GUI -- possibly saving the time of making a mistake. This way a programmer is guided to satisfy constraints as proposed in \cite{burnett_scaling_1995}, unlike syntax errors which just notify the user when he's already breeched one. Finally there is \textbf{the visual overview aspect} -- unlike classical programs where we have to first focus on specific letters and read them to discover what type of algorithm pattern we're dealing with, in visual programming we have the ability to recognize certain patterns in a purely visual way (by recognizing circles, squares, lines, colour etc.), leading to an examination of the meaning of some program in a more top-down manner.

\begin{figure*}[!t]
\centering
\includegraphics[width=7.1in]{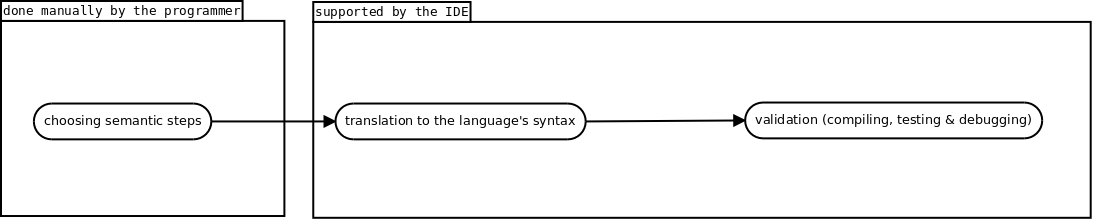}
\caption{A diagram showing the distribution of work between the programmer and the IDE in visual programming}
\label{fig:VPL-diagram}
\end{figure*}

All in all, an attempt is made in VPLs to \textbf{shift more of the work from the programmer to the IDE}. As illustrated in Fig. \ref{fig:VPL-diagram}, the task of syntax generation is done automatically -- allowing the programmer to focus more on the abstract notions of algorithm design.

To summarize, the benefits of this shift are:
\begin{itemize}
  \item \textbf{easy syntax rule obeying} -- only the freedom of creating programs is given to the programmer
  \item \textbf{a better notification system} -- due to the fact that the IDE has more knowledge about the program and the programmer's actions at an earlier moment than classical programming IDE
  \item \textbf{visual representation can be easier to grasp}
\end{itemize}

\subsection{ Related work}

In \cite{burnett_scaling_1995} a general overview of the visual programming strengths and weaknesses is examined with a lot of examples of how certain languages cope with these opportunities and challenges. The two most notable items proposed are:
\begin{itemize}
  \item \textbf{static representation} -- is a graphical notation used to present a program at rest sufficient to understand the logic
  \item \textbf{effective use of computer display} -- showing only the information important to the user at any given time
\end{itemize}

LabView and Simulink can be mentioned as the representatives of a big group -- the data flow VPLs, described in \cite{johnston_advances_2004}, which are based around a functional programming style. A similar VPL, but presented as an online web service is Yahoo Pipes \cite{loton_working_2008}.

KTechlab \cite{_ktechlab_????} is a flowchart VPL used to describe hardware components. Scratch \cite{_scratch_????} is a good example of a WHILE language flowchart VPL that has taken the role of education through fun and interactive app programming. One drawback in Scratch is that it doesn't present the user with the source code generated from his flowchart, thus not encouraging gradual transition to classical programming as the user gets more experienced.

\subsection{ Goals}
\label{sec:goals}
The goal of this paper is to present the methods required to create a VPL for editing and executing flowcharts with an additional ability to generate classical source code from them. This would give the user another educational step between Scratch and classical programming, where he would be able to draw a flowchart and generate good quality, readable code from it and execute it. The code readability would let him study and understand the syntax created from his flowcharts.

These methods will be explained on an example of vIDE, a VPL that was built to fulfil the requirements. The motive behind vIDE is to lower the barrier of learning programming for children as well as for other experts who don't know a programming language syntax, but need to implement certain algorithms.

Firstly, a flowchart editor is needed and the problem of synchronizing a graphical representation with a model will be addressed. Secondly, we will cover the problem of generating code in a certain classical programming language from this model.

An interesting aspect of code generation that will be explored is the internal representation of an algorithm in a GOTO manner, the most similar in logic to a flowchart, and its transformation to a WHILE language data structure. The GOTO and the WHILE languages are formally defined in \cite{saabas_compositional_2006}. In short -- the GOTO language uses a GOTO instruction to jump anywhere in the program, while the WHILE language has WHILE and IF instructions to conditionally execute a certain block (zero, one or multiple times) thus adding more structure and making the flow more predictable. It will be shown that the transformation from a GOTO language to a WHILE language representation is a very practical way of building a flowchart VPL because of the difference between a VPL conceptual model and an output classical programming language.

\subsection{ Organisation of the paper}

In section \ref{sec:methods} the methods used to create a VPL are presented. In \ref{sec:system_architecture} a system architecture of vIDE will be explained so that certain modules can be referenced later in a more clear way. The problem of diagram-to-model mapping is discussed in \ref{sec:flowchart_editor}. The motives and mechanics behind GOTO-to-WHILE model-to-model transformation used in vIDE is given in \ref{sec:goto-while}. Finally, the last step of generating code is shortly explained in \ref{sec:model-code}.

In section \ref{sec:results} we examine the capabilities of vIDE -- a simple example of usage, the special cases of constraint satisfaction and a useful feature of knowledge inferring from the model.

Section \ref{sec:conclusion} places the achieved results in the context of related work and the initial objectives and summarizes the advantages of the used techniques and possible improvements.

\begin{figure*}[!t]
\centering
\includegraphics[width=7.1in]{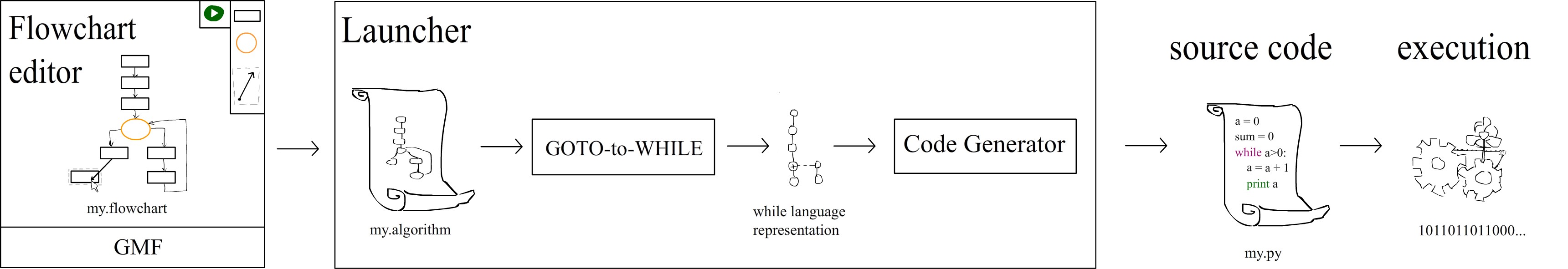}
\caption{A graphical overview of the vIDE system architecture}
\label{fig:vIDE-arhitektura}
\end{figure*}

\section{Methods}
\label{sec:methods}

In order to be fully usable as a programming language, a VPL must allow the programmer to:
\begin{itemize}
  \item express an algorithm's logic in some manner and
  \item be able to translate this logic into executable machine instructions
\end{itemize}

Next, the methods used in implementing our VPL, vIDE, to achieve this will be analyzed.

\subsection{System Architecture}
\label{sec:system_architecture}

The system architecture of vIDE, the VPL used as an example VPL in this paper, is illustrated in Fig. \ref{fig:vIDE-arhitektura}.

In a very general sense vIDE achieves the basic programming language functions through:
\begin{itemize}
  \item a\textit{ flowchart editor} -- where the programmer visually defines his logic
  \item a \textit{launcher} (a compiler of sorts) -- which transforms the user's flowchart into source code that can be executed
\end{itemize}

The first step to achieve this is to allow the user to draw a flowchart. This was implemented using the GMF, a framework for creating diagram editors. As seen in Fig. \ref{fig:vIDE-arhitektura}, an algorithm model is kept in sync with the flowchart that the user draws. GMF does this automatically according to a predefined mapping model). Further information about GMF can be found in \cite{gronback_eclipse_2009}.

The flowchart can be \textit{launched} through the GUI when the user is happy with it. This action initiates a transformation of the algorithm model represented in a GOTO manner to an equivalent WHILE language style representation (this is required because a WHILE representation can't be created by GMF directly from the vIDE flowchart graphical notation -- this will be further discussed in section \ref{sec:goto-while}).

From the WHILE language data structure, a concrete textual syntax can easily generate code in any programming language that relies on WHILE and IF commands to control data flow (for example C \cite{kernighan_c_1988} or Python \cite{_python_????}). In vIDE, a Python syntaxer module is called to generate a Python script as the output program, equivalent to the user's flowchart.

Once the output program is generated, the user can study the source code to learn its syntax or simply run it and observe the effects of the flowchart he drew.

In the next few sections the vital methods needed to create a VPL able to generate code will be examined using vIDE as an example implementation.

\subsection{The flowchart editor -- mapping a diagram to a model}
\label{sec:flowchart_editor}
 
In every VPL some sort of diagram editor is needed that can be map a graphical notation to a model. A flowchart editor in vIDE was created using GMF, which uses a nice abstract way of describing the diagram editor. In this part it doesn't really matter what sort of a diagram we're building (flowcharts are just special types of diagrams). The way GMF works is that it requires several models which define the diagram editor's behaviour. Using these models, the diagram editor Eclipse plug-in can then be generated. Methods for building the flowchart editor will be given here in relation to the GMF architecture, but these or similar modules would be required when creating a diagram editor utilizing any other technology as well.

\begin{figure}[!t]
\centering
\includegraphics[width=3.4in]{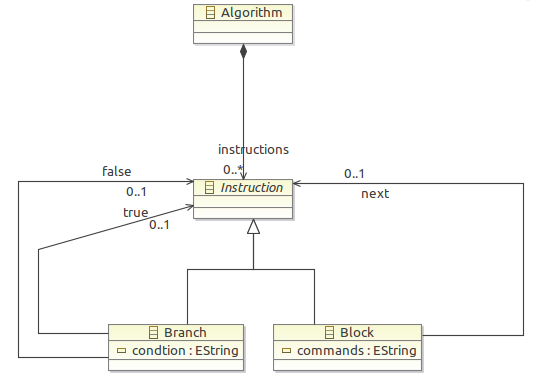}
\caption{Algorithm EMF model used in vIDE - a GOTO-like data structure generated from the flowchart}
\label{fig:vIDE-class_diagram}
\end{figure}

The models needed to create a diagram editor, that is, their implementations for vIDE through GMF are:

\begin{itemize}
  \item \textbf{A graphical definition} -- describing the graphical elements that the user will see in his diagram; in vIDE that would be the block, branch and connection elements.
  \item \textbf{A tooling definition} -- definitions of available tools for drawing the diagram; one tool is defined in vIDE for every graphical element.
  \item \textbf{An Eclipse Modelling Framework (EMF) model} -- this is basically the goal data structure synchronized with the diagram as the user edits it; in vIDE a simple GOTO-like algorithm data structure is used and its class diagram can be seen in Fig. \ref{fig:vIDE-class_diagram}.
  \item \textbf{A mapping model -- }this is the heart of\textbf{ }the\textbf{ }diagram editor description, it defines how to map elements from the graphical definition to an EMF model and GMF uses this definition to synchronize the diagram and the model automatically; if an algorithm model uses a GOTO representation (as is the case in vIDE), this mapping is pretty straightforward -- a block element maps to a block class, branch to branch etc.
\end{itemize}

\subsection{ An algorithm for GOTO to WHILE transformation}
\label{sec:goto-while}
 
After an algorithm model has been created, a program could be generated disambiguously. The problem is that the algorithm model is described in a GOTO language manner and we want to generate a program in a WHILE language. To achieve this, a transformation is necessary. First, we will examine the motives for organizing the data structures in such way that the transformation is necessary and then the transformation algorithm itself.

\subsubsection{Reasons for using a GOTO language}

For a representation of algorithms in flowcharts, a GOTO language seems more natural, because of a single connection (or we might say flow) going out of every block.

Other possible graphical representations were explored - with special graphical elements for WHILE and IF commands, but these approaches were dropped, because there would have to be multiple outgoing flows and additional explanation of these elements' behaviour would be necessary, which would complicate the usage of vIDE.

\subsubsection{Reasons for NOT using a GOTO language}
 
It was explained in the last subsection that GOTO is a suitable language for the graphical representation of an algorithm. For the generated programs, on the other hand, WHILE-languages should generally be used instead.
 
In his open letter, Dijkstra expressed his opinion against using GOTO instructions in programming \cite{dijkstra_go_1979}, while Knuth said in \cite{knuth_structured_1974} that they can be useful at times. Generally, we can conclude that GOTO commands can be used carefully (an example of proper usage of GOTO commands in modern languages are exceptions), but WHILE, IF and similar commands should be used to create structured programs. Structured programs are defined in \cite{knuth_structured_1974}.

One of the good sides of structured programming is readability -- if a programmer wants to see what was generated from his flowchart, it would be much more readable (not to mention educational) if loops would be interpreted as WHILE commands and normal branches as IF commands (instead of everything mapping to simple GOTO commands).

A WHILE language representation of the flowchart might be useful to have in a data structure for visual purposes as well, because the usage of structured programming would allow features such as graph folding (letting the programmer hide unnecessary nodes manually or let the IDE do it for him using context-aware technologies such as \cite{_mylyn_????}).

Python was chosen for an output programming language in vIDE because of its simplicity and readability, so another practical reason for not generating code using GOTO commands in vIDE in specific is that Python doesn't have a GOTO command in its language.

\subsubsection{Combining the two - the constrained GOTO}
\label{sec:constrained_goto}

When a pure GOTO language representation of an algorithm is stored in a data structure it is clearly a \textbf{graph} -- it can contain loops to already executed commands. When an algorithm represented in a structured WHILE programming language is stored it is possible to represent it by a \textbf{tree} -- an intuitive example of this is the ability to fold code in environments such as Eclipse \cite{_folding_????}. The reason this is possible is that the body of a WHILE loop or an IF body can't jump to any other instruction outside its parent WHILE/IF instruction. Even though it's possible to model WHILEs and IFs using GOTO commands, the GOTO would also allow ``forbidden jumps'' (for example to the same command), therefore necessitating a graph data structure for the algorithm's representation.

Since graphs can't generally be represented by trees, to be able to transform a GOTO representation to a WHILE representation, constraints have been introduced in vIDE on the flowcharts that can be drawn in it:
\begin{itemize}
  \item  loops can only be drawn to go back to the last branch predecessor (of the same branching depth)
  \item  if a branch is not a loop its children blocks must join at one point (they need to share a common successor) on the same branching depth
\end{itemize}

Using these two constraints we effectively mask a WHILE language using a GOTO language, without the fear of not being able to convert it to a tree representation.

\subsubsection{The transformation algorithm}
 
Once we set the constraints, we can define an algorithm for transforming a constrained GOTO algorithm (illustrated in Fig. \ref{fig:vIDE-class_diagram}) into a WHILE algorithm.

The GOTO-to-WHILE algorithm is essentially a depth first search (DFS) that recursively processes the nodes of a GOTO algorithm collecting single instructions to blocks and translating branches to WHILE or IF commands using processed node ``colouring'' (see \cite{cormen_introduction_2009} for DFS and similar graph traversal algorithms).

This is the basic idea of the algorithm (meant to be called as G2W(instruction), where the first instruction of the main block of the algorithm should be provided to the initial function call):

\begin{algorithmic}
\WHILE{instruction exists}
	\IF{instruction's a block}
		\STATE nothing special
	\ENDIF
	\IF{instruction's a branch}
    		\STATE trueChild $\gets$ true child of the branch
    		\IF{G2W(trueChild) found processed command}
        		\STATE translate instruction to WHILE
        	\ENDIF
      	\IF{G2W(trueChild) reached program end}
      		\STATE translate instruction to IF
      	\ENDIF
	\ENDIF
	\STATE mark instruction processed
    \STATE instruction $\gets$ next instruction
\ENDWHILE
\end{algorithmic}

 
Note that \textit{G2W(trueChild)} represents processing child instructions of a branch in the case of the condition being met -- going further in depth recursively on the true side. Also, this is only a sketch of the ``interesting'' parts of the algorithm. In an actual implementation:
\begin{itemize}
  \item  All the constraints need to be checked and any breeches communicated to the user. 
  \item  Also, resolving the IF transformation is a bit more complex and requires tracing the condition being false branch as well and fixing the tree afterwards (because at first all the successor instructions would be added to IF as children -- an intersection, i.e. the end of the IF instruction, can only be found after tracing the second branch -- false).
  \item  The last omitted aspect is the communication -- a lot of information needs to be passed as arguments to recursive function calls or using a stack to identify the true state of the graph.
\end{itemize}

These details were skipped, since they would complicate the pseudo-code a lot and they can be implemented quite routinely.

\subsection{ Model-to-code transformation}
 \label{sec:model-code}
 
After we have the algorithm represented in a tree-like structure of a structured WHILE language, the code generation is very simple and consists of:
\begin{enumerate}
  \item  a DFS through the tree
  \item  translating instruction objects to strings in the goal language syntax (Python in vIDE's example) on a one-to-one mapping basis.
\end{enumerate}

\begin{figure}[!t]
\centering
\includegraphics[width=3.4in]{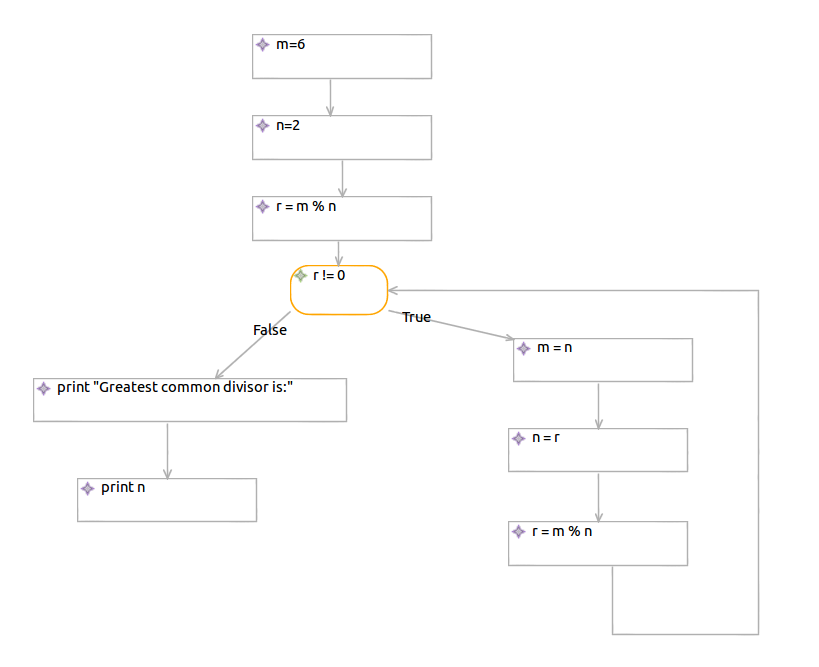}
\caption{A flowchart of the Euclid's algorithm drawn in vIDE}
\label{fig:Euclid-flowchart}
\end{figure}

\section{Results}
\label{sec:results}

As a result of implementing all the presented methods, vIDE was built -- a VPL that can be installed as an Eclipse plug-in and used to draw flowcharts and generate Python code from them. In this section its capabilities will be examined.

\subsection{An example of vIDE usage -- code generation}

The example of the Euclid's algorithm drawn as a flowchart in vIDE is shown in Fig. \ref{fig:Euclid-flowchart}. From this flowchart the user can launch the code generation and a Python script is generated, equivalent to the algorithm defined in the flowchart:

\begin{lstlisting}[language=Python]
m=6
n=2
r = m % n
while r != 0:
    m = n
    n = r
    r = m % n
print "Greatest common divisor is:"
print n
\end{lstlisting}

The branch in the flowchart was recognized by the GOTO-to-WHILE transformation algorithm to be a loop (because an instruction was pointing to its predecessor; note that a constraint wasn't breeched because it is the last branch) so a WHILE instruction in Python was generated.

\subsection{Constraint satisfaction in vIDE}

An example of automatic constraint satisfaction checking is implemented in vIDE, so that the user is not allowed to make a mistake in the first place. For example the user can't connect a block to itself (a circular connection - it would result in an infinite loop), because the environment won't let him ``drop'' a connection on that position. This was implemented using the Object Constraint Language (OCL) \cite{_object_????}, a language for constraint definition which is integrated with GMF in a way that the checks are done in real time -- presenting the user with the currently relevant information.

The constraints defined in section \ref{sec:constrained_goto}) will notify the user about a constraint breech only when he launches the flowchart-to-code transformation. This wasn't implemented in OCL, but in Java (because the check requires recursive function calls, which to our knowledge isn't possible in OCL), as a part of the GOTO-to-WHILE transformation, so the notifications aren't real-time.

\subsection{Knowledge inferring in vIDE}
\label{sec:knowledge_inferring}

One of the nice features of GMF is that it infers knowledge from the EMF model created to describe the diagram editor.

\begin{figure}[!t]
\centering
\includegraphics[width=3.4in]{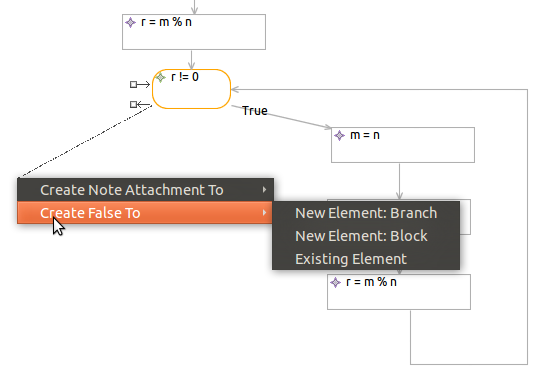}
\caption{Connection creation wth deduction (excess choices are trimmed)}
\label{fig:branch-deduction}
\end{figure}

An example in vIDE is that creating a connection from a branch for the first time asks the user whether it is the true or the false case, but when a true case is already present, the system deduces that there is no need in show this choice to the user as seen in Figure \ref{fig:branch-deduction}, resulting in only the semantically relevant information presented.

\section{Conclusion}
\label{sec:conclusion}

The proposed methods for building a VPL capable of flowchart editing and code generation have been shown to work on the example of vIDE. The VPL can be used for drawing and executing flowcharts comparable to \cite{_scratch_????}, while also giving the user the novel ability of generating clear, readable Python code, achieving the goals set in \ref{sec:goals}. This is important, because it allows the users to more easily grasp programming language syntax and move to more complex, classical programming later (if needed). Such an application could find a lot of use in lowering the programming entry barrier and would be important for:

\begin{itemize}
  \item  didactic purposes -- teaching programming in an interactive and visual way
  \item  ease of usage -- making programming more accessible to people of other professions
\end{itemize}

Features such as knowledge inferring, described in \ref{sec:knowledge_inferring}, provide a great way of leveraging advanced GUI capabilities to provide a better programming environment. This goes in line with the general guidelines for creating VPLs stated in \cite{burnett_scaling_1995} and could be considered as a visual programming equivalent of context aware systems such as Mylyn \cite{_mylyn_????}.

Constraint definition using OCL enables real-time notifications. This seems to be a good way to utilize the advantages of VPLs -- namely a chance for preeminent notifications, as described in \ref{sec:strengths}. One of the problems is checking more complex constraints. Perhaps iterative transformation to the WHILE language model while the flowchart is being drawn would allow for more lively notifications.

\section*{Acknowledgements} 
The authors whish to express their gratitude to Professor Domagoj Jakobović, doc.dr.sc. at the Faculty of Electrical Engineering and Computing, University of Zagreb, for mentoring this research and his help in writing the paper.

 
 \bibliographystyle{IEEEtran}
 \bibliography{kermit-library}
 
\end{document}